\documentclass[runningheads]{llncs}
\usepackage[bookmarks=false]{hyperref} 
\usepackage{graphicx}
\usepackage{hyperref}
\usepackage{rotating}
\usepackage{color, colortbl}
\usepackage{balance}
\usepackage{multirow}
\usepackage{mdframed}
\usepackage{makecell}
\definecolor{Gray}{gray}{0.9}

\usepackage[most]{tcolorbox}
\usepackage{xcolor}
\usepackage{tabularx}
\usepackage{longtable}
\usepackage{orcidlink}

\tcbset {
  base/.style={
    arc=0mm, 
    bottomtitle=0.5mm,
    boxrule=0mm,
    colbacktitle=black!10!white, 
    coltitle=black, 
    fonttitle=\bfseries, 
    left=2.5mm,
    leftrule=1mm,
    right=3.5mm,
    title={#1},
    toptitle=0.75mm, 
  }
}

\definecolor{brandblue}{rgb}{0.34, 0.7, 1}
\newtcolorbox{mainbox}[1]{
  colframe=brandblue, 
  base={#1}
}

\newtcolorbox{subbox}[1]{
  colframe=black!30!white,
  base={#1}
}

\begin{document}
\title{Software Testing Education and Industry Needs - Report from the ENACTEST EU Project}
\titlerunning{Software Testing Education and Industry Needs}
%
\author{Mehrdad Saadatmand\inst{1}\orcidlink{0000-0002-1512-0844} \and
Abbas Khan\inst{1}\orcidlink{0000-0001-6418-9971} \and
Beatriz Marin\inst{2}\orcidlink{0000-0001-8025-0023} \and
Ana C. R. Paiva\inst{3}\orcidlink{0000-0003-3431-8060} \and
Nele Van Asch\inst{4} \and
Graham Moran\inst{5} \and
Felix Cammaerts\inst{6}\orcidlink{0000-0002-0037-3865} \and
Monique Snoeck \inst{6} \orcidlink{0000-0002-3824-3214} \and
Alexandra Mendes\inst{3}\orcidlink{0000-0001-8060-5920}}
%

\authorrunning{M. Saadatmand et al.}
%
\institute{RISE Research Institutes of Sweden, Sweden\\
\email{\{mehrdad.saadatmand, abbas.khan\}@ri.se}\\ 
\and
Universitat Politècnica de València, València, Spain\\
\email{\{bmarin\}@dsic.upv.es}\\
\and
INESC TEC, Faculty of Engineering, University of Porto, Porto, Portugal\\
\email{\{apaiva, afmendes\}@fe.up.pt}
\and
AE nv, Leuven, Belgium\\
\email{\{nele.vanasch\}@ae.be}
\and
NEXO QA, Barcelona, Spain\\
\email{\{graham.moran\}@nexoqa.com}
\and
LIRIS, KU Leuven, Leuven, Belgium\\
\email{\{felix.cammaerts, monique.snoeck\}@kuleuven.be}
}
\maketitle              

\begin{abstract}
The evolving landscape of software development demands that software testers continuously adapt to new tools, practices, and acquire new skills. This study investigates software testing competency needs in industry, identifies knowledge gaps in current testing education, and highlights competencies and gaps not addressed in academic literature. This is done by conducting two focus group sessions and interviews with professionals across diverse domains, including railway industry, healthcare, and software consulting and performing a curated small-scale scoping review. The study instrument, co-designed by members of the ENACTEST project consortium, was developed collaboratively and refined through multiple iterations to ensure comprehensive coverage of industry needs and educational gaps. In particular, by performing a thematic qualitative analysis, we report our findings and observations regarding: professional training methods, challenges in offering training in industry, different ways of evaluating the quality of training, identified knowledge gaps with respect to academic education and industry needs, future needs and trends in testing education, and knowledge transfer methods within companies. 
Finally, the scoping review results confirm knowledge gaps in areas such as AI testing, security testing and soft skills.
\end{abstract}

\section{Introduction}\label{sec:intro}
\setcounter{footnote}{0} 
Given the growing role of software systems and digital services in daily life and industry, coupled with the rapid advancement of technologies and the increasing complexity of software systems, there is a continuous demand for the improvement or development of new, more effective and adaptive software testing techniques and strategies. 

Although there is no doubt that effective software testing is crucial, sometimes it is neglected, which results in flawed and unreliable software applications. The cause is originated from a skills mismatch between what is needed in industry, the learning needs of students and the way testing is currently being taught at Higher Education (HE) and Vocational Education and Training (VET). 

The increasing complexity of software systems (e.g., driven by the proliferation of Internet-of-Things (IoT), adoption of Artificial Intelligence (AI), persistent security vulnerabilities, increasing product variability, and evolving development practices) demands updated software testing courseware that addresses technical challenges, automation needs, security concerns, regulatory compliance, and essential soft skills.

Therefore, to identify and develop software testing teaching materials aligned with industry needs, we need to foster a collaborative environment where HEs, VETs and companies can share knowledge and develop new approaches to software testing education, considering a broader socioeconomic context. 
This will improve students' learning performance and their software testing skills, given that software testing is increasingly important in digital job profiles across the labor market. In the long term, this will improve the quality of the software on which our digitalized society depends.

In this scenario, with all these current and evolving challenges, there is an urgent need for the adaptation of educational programs. 

This paper presents a study performed in the scope of the ENACTEST European project~\footnote{https://enactest-project.eu/}~\cite{marin2022enactest,marin2023enactest} that, by working directly with industry, aims to identify the competences needed by practitioners in their daily jobs and, from there, identify the adaptations needed to be implemented in software testing educational programs.
In this regard, the paper particularly focuses on answering the following research questions (RQs).
\begin{mainbox}{Research Questions} 
    \textbf{RQ1:} \textit{What are the perceived competence needs for testers in industry?}\\
    \textbf{RQ2:} \textit{What are the knowledge gaps in the current testing education?}\\
     \textbf{RQ3:} \textit{Which of the identified industrial competence needs and knowledge\\...\qquad gaps are not covered in the literature?}
\end{mainbox}
Firstly, RQ1 focuses on the collection of data on various trainings and educational sessions provided to testers in professional settings, with the aim of highlighting the needed competences.
RQ1 also focuses on how this knowledge is transferred within the company. Secondly, RQ2 aims to identify the gaps in software testing training and education that need to be addressed. Thirdly, RQ3 focuses on highlighting the unique industrial competence needs that are not covered by existing literature, and in turn, validates the findings of RQ1 and RQ2.

This study investigates industry expectations regarding software testing education in academia, and identifies existing gaps as well as emerging trends.
The identified gaps are useful for enhancing academic curricula and practical training, fostering the alignment with real-world needs. Overall, this work equips the industry with valuable guidance to promote continuous learning in a rapidly evolving body of knowledge.

This paper is structured as follows. Section \ref{sec:enactest} provides a brief description of the European research project, ENACTEST, in which this work is being performed. Section \ref{sec:studydesign} describes the method used to collect qualitative data about software testing industrial needs. Section \ref{sec:results} describes the results and observations of the study performed. Section \ref{sec:validity} presents the threats to validity, and Section \ref{sec:conclusion} concludes and points to future work.

\section{ENACTEST Project}\label{sec:enactest}
ENACTEST \cite{marin2022enactest} is an ERASMUS+ EU project (2022-2025) which addresses the challenges in software testing education from three interconnected perspectives within computer science education: students, industry, and academia. The ENACTEST project aims to create educational materials, known as capsules, which are designed to be concise and easy to incorporate into existing courses without adding extra workload for educators. The capsules introduce innovative teaching techniques in higher education (universities and vocational centers) tailored to students' learning needs and aligned with industry requirements in order to enhance students' competencies and, consequently, their skills as future professionals in the software testing processes.

The project consortium comprises a diverse group of beneficiaries to ensure the outcomes benefit the entire socioeconomic landscape.  The active involvement of industry is crucial for the success and relevance of the ENACTEST results. In response to the rapidly evolving demands of the software sector, the inclusion of small businesses and applied research centers within the ENACTEST consortium plays a key role in ensuring that the educational capsules developed are aligned with current professional practices and industry expectations. Concretely, the ENACTEST consortium includes four universities, one vocational training center, and four small businesses: Universitat Politècnica de València (UPV) - Spain, Katholieke Universiteit Leuven (KU Leuven) - Belgium, Universidade do Porto (UP) - Portugal, Università degli Studi di Napoli Federico II (UNINA) - Italy, RISE Research Institutes of Sweden (RISE) - Sweden, Centro Superior de Formacion Europa-Sur (CESUR) - Spain, NEXO QA -  Spain, INOVA+ - Portugal, and CTG - Belgium.

The capsules undergo a continuous process of refinement and validation \cite{marin2023enactest}. Along with the materials and resources developed, the latest version of each capsule is published on the ENACTEST project website. This enables the broader community to access and utilize them for educational and professional purposes \cite{doorn2025alianza}. 

Moreover, considering the dynamic nature of industry, and the flexibility and adaptability required in research projects \cite{pastor2024industrial}, the ENACTEST team decided to first identify the industry needs directly through relevant industry stakeholders, in order to obtain first-hand information that reflects the most recent demands. After that, the ENACTEST team decided to contrast these needs with the existing literature in order to validate them broadly.

\section{Study Design}\label{sec:studydesign}
In this section, we describe the method used to collect qualitative data about the industrial needs of software testing. 

\begin{figure*}[!th]
      \centering
      \includegraphics[width=1\linewidth]{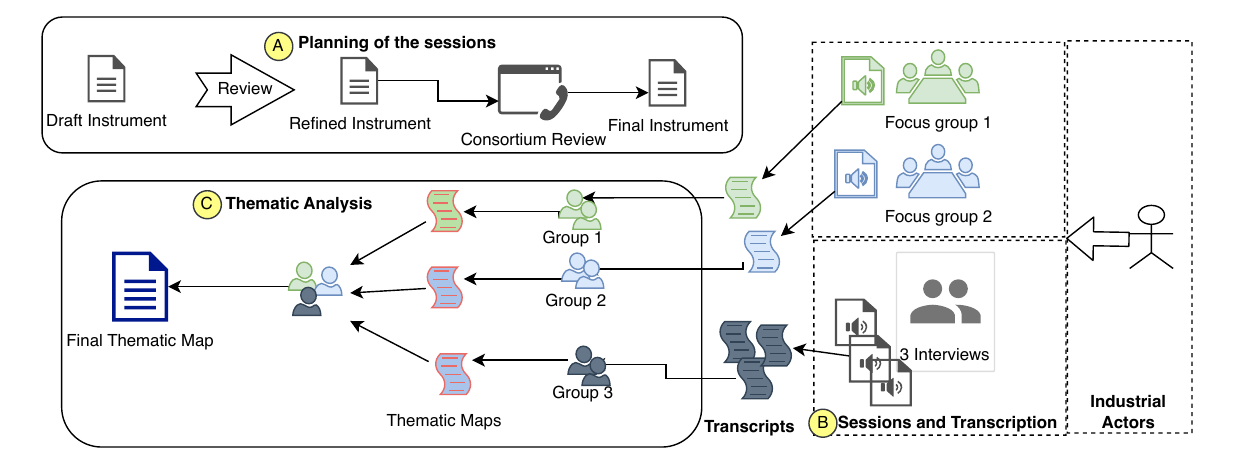}
      \caption{Mixed-method study design to collect qualitative data about industrial needs}
      \label{fig:design}
 \end{figure*}
Figure~\ref{fig:design} illustrates the study design and procedure, which began with planning the sessions and interviews (A). During the planning phase, a series of meetings were conducted to draft an instrument for the study. Then, the instrument was reviewed by the relevant part of the project consortium, and a final version was produced for use in conducting the studies with key actors. This was followed by the session and transcription (B), and thematic analysis (C). Next, we provide a brief overview of the procedure used to obtain and analyze the results. 

\textit{A. Planning of the sessions: } 
As mentioned, an initial instrument was drafted by the ENACTEST partners. Four experts from the consortium (two from industry and two from a research institute) contributed to the development of the initial draft of the instrument. To divide the work better, the instrument was split into three parts, covering data collection for testing, training, knowledge gaps, and knowledge transfer.
The creation of the instrument was guided by the standard terminologies used in the field of software testing.
The produced initial instrument was reviewed, refined, and reordered by all four members of the consortium. 
Finally, the entire ENACTEST consortium was invited to review and refine the instrument, resulting in a final three-part instrument. Below, we briefly summarize the topics covered in the three-part instrument that has been produced. 
\begin{mdframed}[style=style1]
\begin{scriptsize} 
\textbf{Part 1: Software Testing Training }\\
Demographics of software testing training at the participant’s company 
\begin{itemize}
    \item Type (internal or external), and Topic 
\end{itemize}
The mode of Delivery of the training that is given at the participant’s company 
\begin{itemize}
    \item Delivery mode (virtual, on-site), Evaluation method (Exercises, questionnaire, example cases), training material used (slides, case studies, software) 
\end{itemize}
Adequacy of the training to current needs 
\begin{itemize}
    \item Theory vs practice, Interaction, learning evaluation, things missing from the training 
\end{itemize}
\textbf{Part 2: Knowledge Gaps and Void }
\begin{itemize}
    \item Gaps in software testing education, training, business knowledge, tooling, and current practices 
    \item Needed skills from a five-year perspective 
    \item Ways to fill the gaps 
\end{itemize}
\textbf{Part 3: Knowledge Transfer }\\
Knowledge transfer process at the participant’s company 
\begin{itemize}
    \item Mode and direction of transfer (formal or informal), Budget assignment
\end{itemize} 
Mediums of knowledge transfer 
\begin{itemize}
    \item Types of sessions, Tools used, and material used for facilitating knowledge transfer 
\end{itemize}
\end{scriptsize}
\end{mdframed}

In parallel, while drafting the initial instrument, a diverse set of industrial actors was identified, working in a total of six domains. The key industrial actors were informed about the study and asked to provide a list of potential participants that would be relevant to the study. Based on convenience sampling, a total of 12 participants were reached who participated in the two focus group sessions and the interviews performed. We refer readers to Section~\ref{sub:par}-A for more details about the background of the participants. 

Three working groups were formed within the ENACTEST consortium to conduct the sessions and transcribe the recordings.
The working groups were also responsible for analyzing the collected data and reviewing the reported results of the other groups. After the final instrument was drafted (step A) and participants were sampled from the industry (left part of Figure~\ref{fig:design}), the three working groups organized three sessions based on the availability of all the participants. The selected participants who agreed to participate in the sessions were briefed in advance about the ENACTEST project and the session in an email. Two sessions were planned and executed as focus groups, following the guidelines~\cite{breen2006practical}. In addition, the third working group used interviews as a research method to collect data from three interviews. All the sessions used the same study instrument that was produced as output of step (A) of Figure~\ref{fig:design}. In the actual sessions (step B of Figure 1), consent for recording was obtained from the participants, and the sessions were recorded. The sessions lasted from one to two hours and included a brief introduction to the project's objectives and the activity itself. 

The recorded audio files of the session were transcribed and anonymized to omit any confidential information. The transcription of three interviews and two focus groups resulted in a total of 70 pages of Word documents that were again subjected to anonymization of confidential information. The resultant documents were analysed following the guidelines for the commonly used analysis method for qualitative data, thematic analysis~\cite{braun2006using}. 

\textit{C. Thematic Analysis  }
Qualitative data can be analyzed in several different ways. In our case, we use the commonly applied thematic analysis approach following the guidelines proposed by Braun and Clarke~\cite{braun2006using}. We introduce some important background concepts related to thematic analysis as follows. 
\begin{itemize}
    \item \textit{Theme} is an abstraction of a commonly occurring pattern within qualitative data.  
    \item \textit{Sub-Theme} is an abstraction of a sub-pattern within a theme.  
    \item \textit{Codes} act as labels assigned to chunks of qualitative data (such as sentences) for indexing. 
    \item \textit{Thematic Map} is a visual or tabular representation of the extracted themes, sub-themes, and codes. 
\end{itemize}

As mentioned, three working groups were formed by the ENACTEST team that took each session's transcript for analysis. Following the cited guidelines, each transcript was independently coded by each group. The codes were directly extrapolated from the transcripts by identifying keywords that were mentioned during the focus group sessions. For example, the code 'security testing' was identified from the comment "security testing is a necessary skill that modern testers should have". The coded transcripts were then analysed for recurring themes and subthemes. As a result, three thematic maps were derived from the two focus group sessions and the interviews. Members of the working groups reviewed each other's coded transcripts and thematic maps. Then, three working groups collected all the extracted themes and sub-themes in a final collective thematic map where the themes are ranked based on the frequency of occurrence in the transcripts. The resultant thematic map comprises more than 60 themes and sub-themes, each with its corresponding frequency. In the following sections, we present the results in a traceable manner where the themes are linked to sub-themes and codes and are supported by additional evidence (quotes) from the actual transcripts. Note that the quotes from the transcript (in italics) are rewritten for better readability, and context is added in [squared brackets] when needed. 

\subsection*{Participants}\label{sub:par}
\textit{Focus group 1: } The first set of participants for the first focus group were selected based on convenience. To enable diversity in data collected, the domain chosen for the first target group was the safety-critical transportation domain of the railway industry. A company in Sweden was approached to participate in the study and was briefed about the ENACTEST project. We managed to recruit five participants from the railway industry working in two different teams. We ensured diversity among the selected participants in their roles, experiences, and gender. Nevertheless, every participant worked (directly or indirectly) in software testing, verification, and safety compliance in their daily work, with experience ranging from 1 to 15 years.  

\textit{Focus group 2: } The second set of participants --- for a second focus group --- were recruited from the web/mobile application development domain. For this set of participants, we recruited experienced developers and testers based on convenience, with experience ranging from 8 to 15 years. Below, we provide a brief summary of the participants' background. 
\begin{itemize}
    \item Participant 1: 15 years of experience in testing within various sectors, performing manual testing, test automation, quality control for certification, and external and internal audits. 
    \item Participant 2: 10+ years of QA experience within various sectors, focusing on manual testing, test automation for web and mobile applications.
    \item Participant 3: 8 years of QA experience, mainly working with testing automation for web and other applications. 
    \item Participant 4: 10 years of QA experience, focusing on both manual and automated testing for web, mobile, and backend applications.  
\end{itemize}

\textit{Interviews: }
Interviewees 1 and 2: Both representing Partena Professional, brought a combined view from both a technical and operational standpoint. Their experiences added depth to the discussions, especially in terms of integrating technical expertise with business objectives. They currently focus primarily on optimization and standardization within company projects. 

Interviewee 3, from Robovision, provided a unique viewpoint based on their role in overseeing software testing processes. Their contributions shed light on the managerial and strategic aspects of software testing and training. Interviewee 2 has over 20 years of experience in Test Management and Test Automation. As Test Program manager, Interviewee 2 coordinated several programs where multiple agile test teams were working together (in parallel) to deliver a program in the same release. 

Interviewee 4, Representing Daedalus, provided insights from their experience in software development and testing. Their perspective was particularly valuable in understanding the practical applications of testing methodologies within a corporate setting. With more than 15 years of experience in the healthcare industry, they gave insights into how testing is done in their company.

\section{Industry Needs: Thematic Analysis and Observations}\label{sec:results}
In this study, we have consciously deviated from the traditional approach of initiating our research with a literature study. Instead, we began with an industry-driven exploration, positioning real-world practice and current challenges as the foundation for identifying training needs in software testing. The goal was to start from the perspective of what students will need in the work field.

Therefore, we designed and conducted two focus group sessions and a series of in-depth interviews with industrial actors. These participants came from sectors such as the railway industry, software consulting and healthcare. The study instrument was co-developed by different partners and refined collaboratively across the consortium. It focused on gathering data about training types, challenges, knowledge gaps, transfer mechanisms, and future needs. This allowed creating a grounded picture of how testing knowledge is developed and shared in industry settings.

 \paragraph{Professional training (RQ1)}
Across the various interviews and focus groups, it became clear that most companies rely on informal, hands-on approaches to training. Mentoring and learning through real project work stood out as the preferred form of skill development, especially for junior testers. While internal sessions and workshops are common (1), formal external training is used more selectively (2), often as a follow-up after employees have settled into their roles. Pre-recorded online courses (e.g., via Coursera or LinkedIn Learning) are available in some organizations, but participants often found them less engaging and disconnected from their day-to-day work. This can be distilled from the following quote from focus group 2: \textit{``So, the learning platforms that we use the most are Coursera, Udemy, LinkedIn Learning, this is like the three business licences that the company has.''}   One company lets their employees spend a lot of training on how to use Azure DevOPs and make a test plan in it (3).

\paragraph{Challenges }
Several challenges were identified from the focus groups. (1) It's difficult to get all trainings standardized as people have varying degrees of experience.  This feedback was received from our interviews: \textit{``It}\textit{ can take up to 6 months to }\textit{get people up and running on the technical side of testing and the business side of testing. Where }\textit{of course}\textit{ the functional part is the most difficult.}\textit{''} (2) The current training frameworks are not unified and structured enough to standardize skill levels across teams. (3) Balancing the technical aspects of software testing with an understanding of business processes and objectives remains a challenge as well.

\paragraph{Evaluation of the training} 
Participants across interviews and focus groups highlighted differing approaches to training evaluation and ongoing support. Focus Group 1 participants described an informal evaluation process, where the effectiveness of training was observed through project performance and feedback from mentors or line managers. 
In contrast, in Focus Group 2, participants noted that formal evaluations of training in day-to-day testing activities were lacking. They expressed a need for a centralized training portal that would provide continuous access to materials, allow for regular knowledge assessments, and enable timely support from trainers. One participant suggested, \textit{``I would say it could be really useful to have like a space where questions can be asked and at some point during a short period of time you will be answered''}.  
Interview findings echoed this diversity, noting that practices vary across companies. While the presence of a mentor providing ongoing feedback and suggestions was seen as a central component of continuous learning, one participant also emphasized the importance of formal certification, stating that obtaining the ISTQB Foundation certification serves as an objective within the company and adds value to the learning process. 

Together, these perspectives indicate a need for more structured, consistent approaches to both the evaluation and support of training in practice.

 \paragraph{Knowledge gaps (RQ2)}
A number of knowledge gaps were identified too, both for new graduates and in more experienced testers. Key areas that are identified are AI in testing, security testing, and the use of automated tools. Participants and interviewees noted that while theory is sometimes covered in education, practical applications are often forgotten and therefore not developed as a skill. There is also a strong demand for testers to better understand quality standards, domain-specific tools, and how to think like a tester rather than a developer. This is evidenced by following quote from focus group 2, \textit{``Because if you work both as a programmer and as a tester in courses, then you don't really want to find faults in your code.''}. In several cases, the lack of experience with real-life testing cases was seen as a critical shortcoming in existing training programs. Given the popularity of Agile Software Development, specific gaps on Agile testing, Test-driven Development and Behaviour-Driven Development were mentioned frequently. 

\paragraph{Future needs and trends}
Looking ahead, several trends are shaping how training needs are evolving. Security testing is becoming a must-have skill (1). The growing importance of AI and automation was also mentioned frequently (2); not just in terms of testing AI systems, but also in using AI to support test generation, review, and fault detection. This was also mentioned by a participant from focus group 2: \textit{``AI is developing so fast in the technology sector, and we need to keep growing in parallel with it''}.  Beyond technical skills, there’s increasing awareness that testers need a well-rounded profile: communication skills, business understanding, and product knowledge are all seen as essential to succeed in today’s teams (3).

\paragraph{Knowledge transfer}
When it comes to transferring knowledge within teams, most companies blend formal and informal methods. Some teams organize regular sessions—such as weekly QA meetings or monthly deep-dives on specific topics—to keep knowledge flowing and shared across teams (1).  This is a direct answer from a participant in focus group 1: ``\textit{The best way to transfer knowledge is direct communication in one-to-one sessions.}''  One-on-one mentoring is especially important, often supported by documentation, internal wikis, or collaborative tools like Confluence (2). Formal handovers (3) are also used in some organizations when project responsibilities shift.

\vspace{0.4cm}

\begin{figure*}[!htb]
      \centering
      \includegraphics[width=\textwidth]{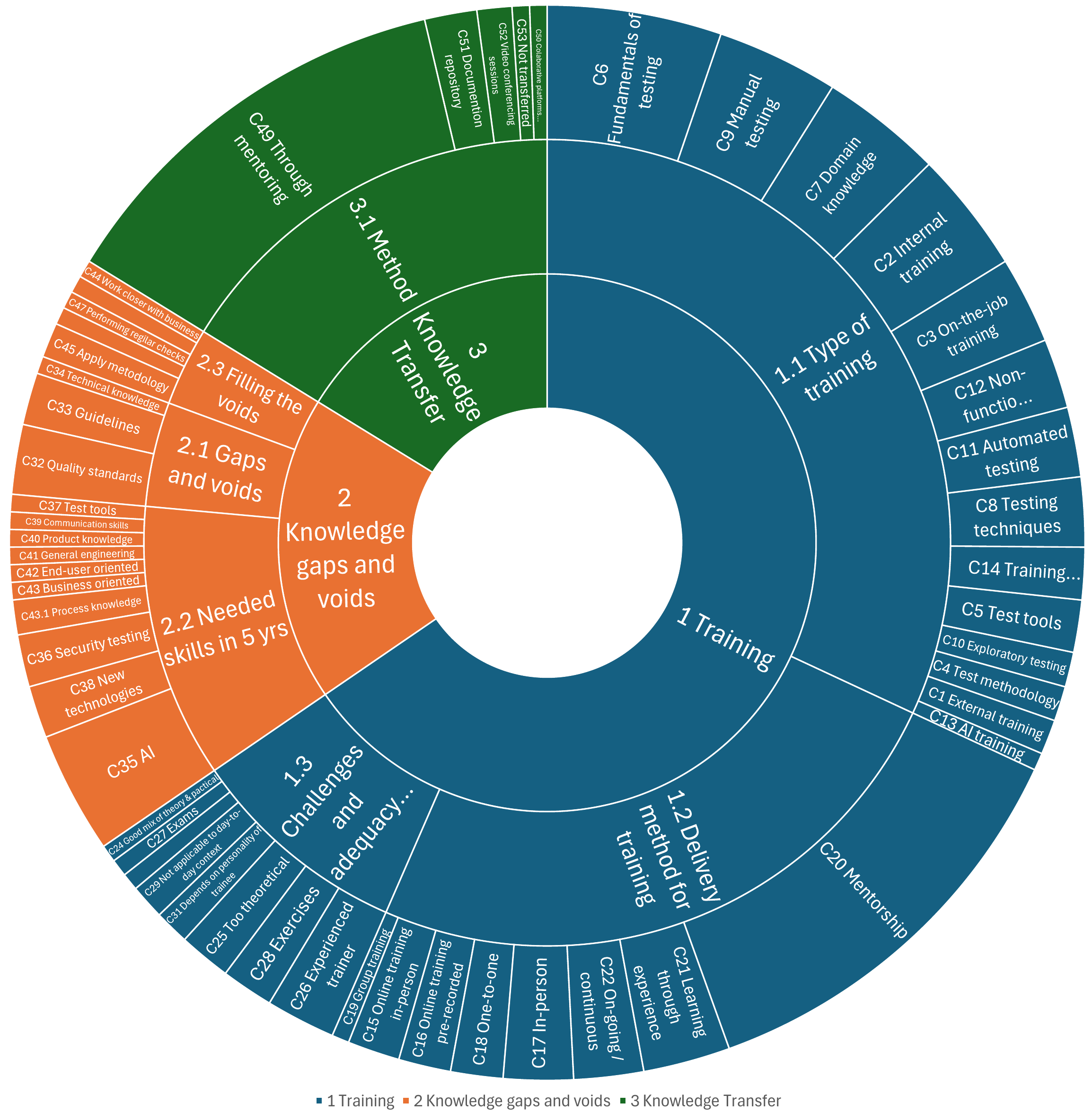}
      \caption{Classification of training needs and practices - themes and codes represented based on frequency}
      \label{fig:codesfrequencydiagram}
 \end{figure*}

As described in Section \ref{sec:studydesign}, to identify and extract the patterns and commonalities specific to training needs, gaps, and knowledge transfer, we classified the gathered industrial inputs by analzying the transcripts from the focus group studies and interviews using different codes and sub-codes. This process helps to represent and highlight specific topics and subtopics discussed within the focus groups and interviews. Considering the appearance frequency of each code topic in the transcripts of the studies, the information is visualized in Figure \ref{fig:codesfrequencydiagram}. A detailed analysis of this information is provided in Deliverable 3.1 of the ENACTEST project~\footnote{https://enactest-project.eu/resources/project-results/}.

\normalsize

\subsection{Discussion and Reflections}
In this section, we discuss the most prominent themes and observations across the studies with respect to the demographics of training in industry in the light of the identified needs, and also present a summary of the testing needs, trends and knowledge transfer methods. 

Our analysis across the sessions with key industrial actors shows that the onboarding process of these companies often includes sharing of training materials with new employees, briefing them about the way of working, standards and processes. This is very well in line with the training needs we identified about processes and standards. The collected data shows a trend that most new employees again have to learn new processes and standards when employed at a new company. This shows the lack of training and education on standard ways of working (for example, the gated process followed in safety-critical systems or test-driven development). The analysis further indicates that trainings are then requested as the employees go forward in their roles. Trainings are also often internal; nevertheless, a limited number of external trainings are also used to train the employees in a specific problem area or tooling. A very prominent and effective model of training was found to be the application of mentorship supplemented with learning through experience. Many of the actors use a senior developer or tester to train junior employees in real-world projects. This seems to be effective, as new employees get their hands on real projects and still have a learning environment created by a senior mentor. In general, the following testing training categories were identified: i) training during on-boarding, ii) training based on special requests, iii), internal vs external, and online vs on-site training, and iv) mentorship and learning through experience. 


Certain challenges regarding software testing training were also identified in our analysis. In particular, our analysis shows that non-interactive online trainings (often pre-recorded) lack the human touch and is often not as effective. In addition, at times, employers expect employees to find the time outside working hours to complete such training. In this regard, there is a need for more on-site and interactive software testing training to address the current need in the market.
Another challenge we observed was the lack of balance between theory and practice, and the lack of material on new/emerging technologies for employees. It can also happen that the content of the training is often too disconnected from the daily work of the employees and does not specifically address their actual needs. Trainings can also often not be based on real cases for demonstration and training materials, making training less relevant to the daily needs of employees. Moreover, training materials could get outdated, and thus not account for new approaches and advances within software testing. This is also very well aligned with the identified needs, where, for instance, security testing, testing of AI-based systems, and the use of AI in testing were highlighted among the many identified needs in our industrial actors' inputs. In addition, prominent patterns in data from industrial actors also show the need for domain-specific knowledge, processes (agile and test-driven development) and knowledge about domain-specific standards. 

With respect to knowledge transfer in industrial organizations, our study highlights the use of centralised repositories and recorded materials to transfer and disseminate knowledge within the organizations and testing teams. Furthermore, existing documentation and project history are also used to transfer knowledge in smaller teams (for example, specific project teams). Active knowledge is transferred either with formal handover sessions for knowledge transfer, one-to-one sessions, or collaborative workshops.

\subsection{Validation}
To validate these findings, we performed a curated small-scale scoping review \cite{levac2010scoping}. In this scoping review, we aim to identify which industry needs have already been addressed in academic literature. After all, there is no guarantee that an identified need has never been addressed since academic research does not consistently get integrated into practice \cite{graham2006lost,ioannidis2016most}.

The search query constructed for our scoping review utilized the following logical combinations of keywords:
\noindent(\texttt{``software tester skills''} $\vee$ \texttt{``quality assurance skills''} $\vee$ \texttt{``QA skills''}) $\wedge$ (\texttt{``future of software testing''} $\vee$ \texttt{``future skills software testing''}).

We used this search query on several platforms: SCOPUS and Google Scholar for scientific literature, and industry-specific sources such as Capgemini, Practitest, and TestSigma for industrial reports. The search query yielded a total of 43 papers. 

We filtered these 43 papers with a predefined set of inclusion and exclusion criteria to ensure the scientific rigor and relevance of the identified reports.

\noindent Inclusion Criteria:
\begin{itemize}
    \item Evidentiary Basis: Studies providing empirical evidence derived from previously peer-reviewed publications or recognized scientific reports.
    \item Methodological Rigor (Survey-Based Research): Primary research reports that utilized survey methodologies with a clearly defined and demonstrably representative sample of participants, ensuring generalizability of findings to the target population.
\end{itemize}

\noindent Exclusion Criteria:
\begin{itemize}
    \item Temporal Relevance: Publications predating the last five years (i.e., prior to 2020), to ensure currency of findings in an evolving domain.
    \item Lack of Empirical or Methodological Foundation: Studies lacking a clear articulation of their methodology, empirical data, or theoretical framework (e.g. opinion papers), thus precluding an assessment of their validity or reliability.
\end{itemize}

Applying the inclusion and exclusion criteria left us with 30 papers that we considered relevant.
We augmented this selection of 30 papers by performing backward snowballing: references from the selected papers were systematically checked for additional relevant publications (1-step backward snowballing) based on title. Each identified paper was rigorously assessed against the predefined keywords and the detailed inclusion and exclusion criteria described above to determine its qualification for inclusion in our review. Duplicates were removed.

After the comprehensive search and snowballing, a compiled list of 71 reports was established, comprising 43 reports from the initial search and an additional 28 papers identified through snowballing. \footnote{All the identified papers can be found at \url{https://anonymous.4open.science/r/IndustryNeeds-6C98/}}

We coded the information present in the papers concerning common trends and future skills required for software testing. More specifically, we aimed to gather following information: empirical validation, representative domain, identified needs, identified solutions for needs, identified knowledge transfer possibilities, identified tools for knowledge transfer, identified risks.

We then compared the identified codes of the scoping review with the prior identified codes of the focus groups and interviews (Section \ref{sec:results}). 

This yielded us following list of topics that were identified in the focus group, but not identified in our scoping review:

\begin{enumerate}
    \item Training
    \begin{enumerate}
        \item Type of Training
        \begin{enumerate}
            \item Manual testing: hands-on execution of test cases without automation
            \item Automated testing: using scripts and tools to run test cases
            \item Non-functional testing: testing for performance, security and usability
        \end{enumerate}
        \item Delivery method of Training
        \begin{enumerate}
            \item Online training pre-recorded: self-paced online courses 
            \item In-person: instructor-led virtual learning
            \item One-to-one: individual coaching or training
            \item On-going / continuous: continuous professional development
            \item Lunch sessions: informal learning discussions
        \end{enumerate}
        \item Challenges and adequacy of training
        \begin{enumerate}
            \item Too theoretical: training lacks real-world application
            \item Exams: assessing knowledge through tests
            \item Not applicable to day-to-day context: training does not match real work needs
        \end{enumerate}
    \end{enumerate}
    \item Knowledge gaps and voids 
    \begin{enumerate}
        \item Gaps and voids
        \begin{enumerate}
            \item Quality standards: lack of clarity in quality benchmarks
            \item Guidelines: absence of clear testing protocols
        \end{enumerate}
        \item Needed skills in 5 years
        \begin{enumerate}
            \item End-user oriented: focusing on user needs
        \end{enumerate}
        \item Filling the voids
        \begin{enumerate}
            \item Apply methodology
        \end{enumerate}
    \end{enumerate}
    \item Knowledge Transfer 
    \begin{enumerate}
        \item Method
        \begin{enumerate}
            \item Documentation repository: sharing knowledge via digital tools
            \item Not transferred: knowledge retention issues
        \end{enumerate}
    \end{enumerate}
\end{enumerate}

Consequently, our analysis of these 71 reports revealed that while current industry practices prioritize hands-on, practical training methods such as mentoring and on-the-job learning, there are still significant gaps not covered by research, especially in emerging areas like AI testing, security testing, and integrating new technologies. Soft skills, communication, and a stronger connection between testing practices and business goals also emerged as critical areas for development.
\section{Validity Threats}\label{sec:validity}
While this study was carefully designed and executed to explore the training needs and knowledge gaps in software testing for industry needs, we analyze the threats to validity based on \cite {wohlin2012experimentation}. 

\paragraph{Internal Validity} The presence of researcher bias during the thematic analysis or in the interpretation of data cannot be excluded. However, to minimize it, we organized three independent working groups to conduct the coding of the data transcripts. Each group performed coding independently, after which cross-reviewing of coded transcripts and thematic maps were collected. This process helps ensure consistency and robustness in the final identified themes. Also, the instrument was carefully designed, with several people involved, to also reduce bias such as the Hawthorne effect.

\paragraph{Construct Validity} of focus groups can be threatened by factors such as ambiguous construct definitions, where participants may interpret 'testing needs' differently; social desirability bias, where individuals adapt their answers to fit the group or please the moderator; and moderator influence, where leading questions can shape the discussion in unintended ways. To mitigate these risks, researchers have developed a protocol involving clear operational definitions of constructs, the use of neutral and well-piloted questions, moderator training to minimise bias and encourage diverse viewpoints, and triangulation with other data sources (e.g. individual interviews) to validate the accuracy of the group discussion in reflecting the intended construct.

\paragraph{External Validity} The relatively small and convenience-based sample of participants may not be representative of the broader software testing industry, limiting the generalizability of the findings. However, the sample includes participants with diverse roles and experience, and we validated our findings with the literature. This study was aimed to be qualitative, not quantitative, to obtain contextual insights into current industry practices. In addition, although most of the participants in Focus Group 2 were seniors
, which could skew views toward more experienced perspectives, Focus Group 1 included less experienced participants
, which adds some balance to the experiment.

\section{Conclusion and Future Work}\label{sec:conclusion}
This paper has presented a study conducted within the European ENACTEST project, which aims to develop educational materials tailored to the evolving demands of software testing education. 

The main goal of this study was to identify both current and emerging training needs in software testing, focusing on the perspective of what students will need in the real world. 
It was based on two focus group sessions, a series of interviews with professionals from different sectors and a curated literature review. 
By combining these research methods, it was possible to validate the findings,  identify overlapping themes, pinpoint unique needs emerging from each approach, and identify major trends and common needs for future training.

This work revealed that while current industry practices prioritise hands-on, practical training methods such as mentoring and on-the-job learning, there are still significant gaps, especially in emerging areas like AI testing, security testing, and integrating new technologies. Soft skills, communication, and a stronger connection between testing practices and business goals also emerged as critical areas for development. These results are relevant for practitioners and researchers, who can use this information to provide students and novice practitioners with tailored training to help them acquire testing competence more quickly.

Moreover, the results obtained were instrumental for the design of educational materials that address identified educational gaps, respond to evolving industry requirements, and promote pedagogical effectiveness. More concretely, using these results we developed 17 diverse teaching capsules that support the industry needs and focus on different testing topics, such as GAMFLEW \cite{silva2025gamflew} that supports the industry need \textit{Interactive sessions ensure real-time engagement and feedback}, ModelDefenders \cite{cammaerts2024modeldefenders} which supports the industry need \textit{Group training fosters collaborative problem-solving}, TSGame \cite{fasolino2024test} that supports the industry need \textit{Clear guidelines improve test planning and execution}, or GADGETS \cite{doorndesign} that supports the industry need \textit{Exploratory testing helps uncover hidden defects}.

The teaching capsules~\footnote{https://enactest-project.eu/capsules-repository/} developed in the ENACTEST project cover 79\% of the needs identified in this work, demonstrating their relevance and identifying opportunities for future work, mainly addressing underrepresented areas. Nevertheless, it is important to note that there exist industry needs that can not be covered by using capsules of education as they must be experienced in live environments, for instance the lunch sessions for informal learning of exploratory testing.

Finally, we advocate that the results obtained equip the industry with guidance to promote continuous learning in a constantly evolving area. Future work considers to implement further capsules to support the remaining industry needs as well as a broader application of capsules at different companies, and higher education institutions.


\begin{credits}
\subsubsection{\ackname} This work has been partially funded by ENACTEST (European
innovation alliance for testing education) ERASMUS+ Project number 101055874, 2022-2025. The work is also supported by the Swedish Knowledge Foundation (KKS) through the ARRAY project. We would also like to thank the industry experts and professionals who participated in our studies and provided valuable inputs. 

\end{credits}
%
%
%

 \bibliographystyle{splncs04}
 \bibliography{references}

\end{document}